\documentclass[12pt,preprint]{aastex}

\received{}
\accepted{}

\begin{document}

\title{The Effect of Radiative Cooling on the Sunyaev-Zel'dovich
       Cluster Counts and Angular Power Spectrum: Analytic Treatment}

\author{Yu-Ying Zhang and Xiang-Ping Wu}

\affil{National Astronomical Observatories, Chinese Academy
                 of Sciences, Beijing 100012, China}

\begin{abstract}
Recently, the entropy excess detected in the central cores of groups 
and clusters has been successfully interpreted as being due to 
radiative cooling of the hot intragroup/intracluster gas.
In such a scenario, the entropy floors $S_{\rm floor}$ 
in groups/clusters at any given redshift are completely determined 
by the conservation of energy. In combination with 
the equation of hydrostatic equilibrium and the universal density profile 
for dark matter, this allows us to derive the remaining gas distribution 
of groups and clusters after the cooled material is removed. 
Together with the Press-Schechter mass function we are 
able to evaluate effectively how radiative cooling can modify 
the predictions of SZ cluster counts and power spectrum. 
It appears that our analytic results are in good agreement with
those found by hydrodynamical simulations. Namely, cooling leads to
a moderate decrease of the predicted SZ cluster counts and 
power spectrum as compared with standard scenario. However,
without taking into account energy feedback from star formation which 
may greatly suppress cooling efficiency, it is still premature
to claim that this modification is significant for the cosmological
applications of cluster SZ effect.  
\end{abstract}


\section{Introduction}

With the rapid development of observing techniques
(Birkinshaw, Hughes \& Arnaud 1991; Jones et al. 1993;
Birkinshaw \& Hughes 1994; Carlstrom,  Joy 
\& Grego 1996; Myers et al. 1997; Hughes \& Birkinshaw 1998; etc.), 
the Sunyaev-Zel'dovich (SZ) effect has become one of the most 
powerful tools for the detections of high redshift clusters
(Joy et al. 2001) and cosmic microwave background (CMB) 
anisotropy on small scales (Mason et al. 2002).  
Indeed, the redshift-independence is the major 
advantage of non-targeted SZ surveys over traditional 
optical and X-ray observations. This arises from the fact that
the SZ effect depends uniquely on the intrinsic properties of 
the warm-hot gas associated with cosmic structures, whilst the photons 
interacting with the gas come from CMB at very high 
redshift $z\approx1000$. 
Because robust constraints on cosmological models are provided
by the most massive and distant clusters, 
growing interest over the past years has been focused upon how
well the fundamental cosmological parameters can be constrained
with non-targeted SZ cluster surveys (Molnar, Birkinshaw \& Mushotzky 2002)
and SZ power spectrum (Bond et al. 2002).
The sensitivity of the expected SZ cluster counts 
and SZ power spectrum to the 
underlying cosmological model is quite impressive.

On arcminute scales, the strength of thermal SZ signals is
directly proportional to 
the total thermal energy of the hot gas confined
in clusters. Unlike the dark matter component of clusters whose
dynamical behavior is governed purely by gravity, the 
intracluster gas is easily disturbed by many complicated physical 
processes in addition to gravity and thermal pressure, which include
(non)gravitational heating, radiative cooling, star formation 
and energy feedback, magnetic fields, etc. 
The reliability of cosmological applications of 
SZ surveys on arcminute scales is closely connected with the question of 
how well one can handle these non-gravitational mechanisms. At present, 
a sophisticated treatment of the problem must rely on   
hydrodynamical simulations coupled with some semi-analytic 
approximations (e.g. da Silva et al. 2000, 2001; 
Seljak, Burwell \& Pen 2001; Zhang, Pen \& Wang 2002; 
White et al. 2002). 
Yet, the effective probe of cosmological
models using non-targeted SZ surveys can be more easily achieved
by semi-analytic approaches because of the requirement
of the continuity in cosmological parameter space.
It is thus desirable to understand the essential physics
that dominates the dynamical evolution of intracluster gas.

While intracluster gas is mainly driven by gravitational shocks
and adiabatic compression, 
two completely different mechanisms have been suggested thus far
that may significantly affect the distribution of the intracluster gas:
nongravitational preheating and radiative cooling.
This is primarily motivated by the steepening of the
X-ray luminosity-temperature relation and the entropy excess of 
groups and clusters reported by a number of investigators
over the past years (e.g. David et al. 1993; Wu, Xue \& Fang 1999; 
Ponman, Cannon \& Navarro 1999).
However, it has been realized that 
the prevailing preheating scenario is facing the so-called
energy crisis -- an unreasonably high efficiency of
energy injection into the intracluster medium from
supernovae must be required (Wu, Fabian \& Nulsen 2000). 
Moreover, 
a uniform preheating of the cosmic baryons to a temperature of
$\sim10^6$ K would make the Ly$\alpha$ forest to disappear
at high redshifts. In contrast, radiative cooling is a natural 
process of the hot intracluster gas, in which no exotic physics 
is needed. Several recent studies have shown that radiative 
cooling alone can allow one to successfully reproduce not only 
the observed distributions of global X-ray luminosity and entropy 
but also the internal structures of hot gas in groups and clusters
(Bryan 2000; Pearce et al. 2000; 
Muanwong et al. 2001, 2002; Voit \& Bryan 2001; Wu \& Xue 2002a,b; 
Borgani et al. 2002; Voit et al. 2002). 
If confirmed, this would have profound implications for 
our understanding of the evolution of hot gas and the formation
of stars in the most massive systems in the universe.

So far, only can one employ hydrodynamical simulations to
assess the effect of radiative cooling on the SZ counts and power
spectrum (da Silva et al. 2000; White et al. 2002), 
while in the preheating scenario
several (semi)analytic models have been 
applied for the prediction of SZ cluster surveys  
(Cavaliere \& Menci 2001;
Holder \& Carlstrom 2001; Benson, Reichardt \& Kamionkowski 2002).
Now, a fully analytic treatment of radiative cooling effect on the 
prediction of SZ cluster counts and power spectrum
may become possible if the observed central entropy floor
of groups and clusters can be attributed to 
radiative cooling of hot intracluster/intragroup gas 
(Voit \& Brayn 2001). In such a scenario,
the minimum entropy of gas distribution in each cluster 
can be uniquely determined by the conservation of energy,
which is equivalent to specifying the equation of state for the 
intracluster gas. Consequently, we will be able to derive
the gas distribution of clusters at various redshifts 
as a result of cooling, in combination
with the hydrostatic equilibrium hypothesis, provided that 
the underlying dark matter profile can be approximated by some
kinds of analytic form, e.g., the  universal density profile  
(Navarro, Frenk \& White 1997; NFW). 
Together with the Press-Schechter (1994; PS) formalism for 
the abundance of dark halos at different cosmic epoch, 
we can eventually compare the predicted SZ  
counts and power spectra with and without radiative cooling, 
and demonstrate the uncertainty 
in the determination of the cosmological parameters arising from
radiative cooling.

\section{Gas distribution with and without cooling}

\subsection{Dark matter distribution}

Intracluster gas with and without cooling
is always assumed to be in hydrostatic
equilibrium with the underlying 
gravitational potential dominated by dark matter component $\rho_{\rm DM}$:
\begin{equation}
\frac{1}{\mu m_{\rm p} n_{\rm e}}\frac{d(n_{\rm e} k_{\rm B} T)}{dr}=
  -\frac{GM_{\rm DM}(r)}{r^2}, 
\end{equation}
where $n_{\rm e}$ and $T$ are the electron density and temperature,
respectively, and $\mu=0.585$ is the mean molecular weight.
We use the universal density profile suggested by
numerical simulations (NFW) for $\rho_{\rm DM}$
\begin{equation}
\rho_{\rm DM}(r)=\frac{\delta_c\rho_{\rm crit}}
                 {(r/r_s)(1+r/r_s)^2}, 
\end{equation}
where $\delta_c$ and $r_s$ are the 
characteristic density and length of the halo, respectively,
and  $\rho_{\rm crit}$ is the critical
density of the universe at cosmic time $t$.
We follow the prescription of Eke, Navarro \& 
Steinmntz (2001) to fix the two free parameters, $\delta_c$ 
and $r_s$, in the NFW profile. To do this, the concentration 
parameter $c=r_{\rm vir}/r_s$ of a dark halo identified at redshift 
$z$  is related to the collapsing redshift $z_{\rm coll}$ through
\begin{equation}
c^3=\frac{\Delta_{\rm c}(z_{\rm coll})}{\Delta_{\rm c}(z)}
  \frac{\Omega_{\rm M}(z)}{\Omega_{\rm M}(z_{\rm coll})}
  \left(\frac{1+z_{\rm coll}}{1+z} \right )^3,
\end{equation}
where $\Delta_{\rm c}$ is the overdensity of dark halo 
within virial radius $r_{\rm vir}$ with respect to
$\rho_{\rm crit}$, for which we take 
$\Delta_{\rm c}=18\pi^2+82[\Omega_{\rm M}(z)-1]-39[\Omega_{\rm M}(z)-1]^2$ 
for a flat universe, and
$\Omega_{\rm M}(z)$ is the cosmic density parameter.
The collapsing redshift $z_{\rm coll}$ is determined by
\begin{equation}
D(z_{\rm coll})\sigma_{\rm eff}(M_s)=\frac{1}{C_{\sigma}},
\end{equation}
where $D$ is the normalized linear growth factor, $C_{\sigma}=25$
for $\Lambda$CDM model, and $\sigma_{\rm eff}(M_s)$ is
the so-called modulated rms linear density at mass scale $M_s$
\begin{equation}
\sigma_{\rm eff}(M_{\rm s})=\sigma(M_{\rm s})
              \left[-\frac{d\ln\sigma(M_{\rm s})}
                          {d\ln M_{\rm s}}\right],
\end{equation}
and $M_s$ corresponds to the mass contained within $r=2.17r_s$ 
where the circular velocity reaches the maximum.
Finally, we specify a temperature $T_{\rm vir}$ 
to the hot gas in cluster
of mass $M$ at redshift $z$ in terms of cosmic virial theorem 
(Bryan \& Norman 1998)
\begin{equation}
k_{\rm B} T_{\rm vir} = 1.39 f_{\rm T}(h^2 \Delta_{\rm c}
 E^2 )^{1/3}\; {\rm keV}
 \left ( \frac {M}{10^{15}M_{\odot}}\right )^{2/3},
\label{eq:tvir}
\end{equation}
in which we will choose the normalization factor to be $f_T=0.92$,
$E^2=\Omega_{\rm M}(1+z)^3+\Omega_{\Lambda}+(1-\Omega_{\rm M}-
\Omega_{\Lambda})(1+z)^2$, 
and $h$ is the Hubble constant in units of  100 km s$^{-1}$ Mpc$^{-1}$.

\subsection{Gas distribution without cooling}

In the absence of cooling, we adopt two models for the gas
properties: 

Model I -- Gas is assumed to follow the same distribution
as the dark matter in clusters:
\begin{equation}
n^0_{\rm e}=\frac{f_{\rm b}}{\mu_{\rm e} m_{\rm p}}\; \rho_{\rm DM}
\end{equation}
in which $f_{\rm b}$ is the universal baryon fraction, and 
$\mu_e=2/(1+X)$ is the mean electron weight with $X=0.768$ being
the hydrogen mass fraction in the primordial abundance of hydrogen 
and helium.  We solve the equation of hydrostatic equilibrium 
to get the temperature profile under the boundary
restriction $T^0(r\rightarrow\infty)\rightarrow0$:
\begin{equation}
k_{\rm B}T^0(r)=k_{\rm B}T^* \frac{r}{r_s}\left(1+\frac{r}{r_s}\right)^2
      \int_{r/r_s}^{\infty}\frac{(1+x)\ln(1+x)-x}{x^3(1+x)^3}dx,
\end{equation}
where $k_{\rm B}T^*=4\pi G \mu m_p \delta_c \rho_{\rm crit} r_s^2$.

Model II -- Gas is assumed to be isothermal and $T^0(r)=T_{\rm vir}$.
In this case, the electron number density in terms of 
the equation of hydrostatic equilibrium  reads (Makino, Sasaki \& Suto 1998)
\begin{equation}
n^0_{\rm e}(r)=n_{\rm e0} e^{-\alpha} 
                \left(1+\frac{r}{r_{\rm s}}\right)^{\alpha/(r/r_{\rm s})},
\label{eq:neiso}
\end{equation}
where $\alpha=4 \pi G \mu m_{\rm p}\delta_c\rho_{\rm crit}
r_{\rm s}^2/k_{\rm B}T$. 
The normalization parameter $n_{\rm e0}$ can be fixed through
\begin{equation}
\int _{0} ^{r_{\rm vir}} 4\pi\mu_{\rm e} m_{\rm p} r^2 n^0_{\rm e}(r) dr =
       M_{\rm vir} f_{\rm b}.
\end{equation}

\subsection{Gas distribution with cooling}

The cooling time scale for a steady, highly subsonic flow 
is determined by the conservation of energy. Setting the energy loss rate
due to bremsstrahlung emission to equal the change in the specific 
energy of gas yields 
\begin{equation}
t_{\rm c}=2.869 \times 10^{10}  {\rm yr}\;\left
( \frac{1.2}{g(T^0)} \right )
\left
(\frac{k_{\rm B} T^0}{\rm keV}
\right )^{1/2}\left (\frac{n^0_{\rm e}}{10^{-3} {\rm cm^{-3}}}
 \right )^{-1},
\label{eq:tcool}
\end{equation}
where $n^0_{\rm e}$ and $T^0$ are the electron number density and
temperature without cooling,
respectively, and $g$ is the total Guant factor. 
If the cooling time $t_c$ is chosen to be the age of clusters, 
or approximately the age of the universe, 
the above equation would allow us to set up a link between 
$n^0_{\rm e}$ and $T^0$. Note that this may lead to an overestimate of
the cooling effect, and meanwhile $t_c$ becomes to be cosmological model 
dependent. Perhaps, a more reasonable approach is to define 
the age of a cluster as the cosmic time between its 
collapsing redshift $z_{\rm coll}$ and the redshift $z$ when it is 
identified (cf. Section 2.1).
Once the gas density is specified, we will be able to work out
the cooling radius and the amount of gas that cools out of the hot
phase by the time $t_c$ (e.g. Wu \& Xue 2002b). 
The fate of the cooled materials within
cooling radius may be associated with star formation. Now, our task is to
work out the new distribution of the intracluster gas with cooling under 
different assumptions of the equation of state.

Model III -- Before and after cooling, the equation of state for gas is 
always assumed to be isothermal and $T(r)=T_{\rm vir}$. 
Consequently, the functional form of equation (\ref{eq:neiso}) also 
applies to the cooling case. However, the normalization after cooling 
is made through
\begin{equation}
\int _{0} ^{r_{\rm vir}} 4 \pi\mu_{\rm e}m_{\rm p} r^2 n_{\rm e}(r) dr =
	     M_{\rm vir} f_{\rm b} - M_{\rm cool}
\end{equation} 
where the total cooled material $M_{\rm cool}$ is given by
\begin{eqnarray}
M_{\rm cool} & = & \int _0 ^{r_{\rm cool}}
                  {4 \pi\mu_{e}m_{\rm p} n_{\rm e}^0 (r) r^2 dr},\\
             & = & 4
                  \pi f_{\rm b} \delta_{\rm c}\rho_{\rm crit} r_{\rm s}^3
                  \left[\ln\left(1+\frac{r_{\rm cool}}{r_{\rm s}}\right)
                  - \frac{r_{\rm cool}}
                  {r_{\rm s}+r_{\rm cool}} \right],
\label{eq:mcool}
\end{eqnarray}
and $r_{\rm cool}$ is the cooling radius which can be
obtained by combining equations (\ref{eq:neiso}) and (\ref{eq:tcool}).

Model IV -- Gas traces dark matter (Model I) before cooling, and the
equation of state for the new gas distribution after cooling is 
given by the entropy before cooling plus a constant entropy floor
$S_c$ (cf. Holder \& Carlstrom 2001; Voit \& Bryan 2001)
\begin{equation}
S=\frac{k_{\rm B}T}{n_{\rm e}^{2/3}}
=S_{\rm c}+\frac{k_{\rm B}T^0}{(n^0_{\rm e})^{2/3}},
\end{equation}
where the entropy floor is determined by the cooling time $t_c$
\begin{eqnarray}
S_c &=&  \frac{k_{\rm B}T_{\rm vir}}{{(n^0_{\rm e}})^{2/3}} \nonumber\\
    &=& 100\; {\rm keV \; cm^2}\;  
 	\left(\frac{t_{\rm c}}{2.869\times 10^{10}{\rm yr}}\right)^{2/3}
	\left(\frac{g}{1.2}\right)^{2/3} 
	\left (\frac{k_{\rm B}T_{\rm vir}}{\rm keV}\right)^{2/3}.
\end{eqnarray}
This ``entropy floor'' increases with cosmic time, resulting in
a deposition of cooled materials in the central regions of clusters
(Voit \& Bryan 2001). Solving the equation of hydrostatic equilibrium
yields
\begin{equation}
k_{\rm B}T(r) = - \frac{2}{5} G \mu m_{\rm p} S^{3/5}(r)
                       \int_{r} ^{\infty}
                       \frac{M_{\rm DM} (r)}{r^2} S^{-3/5}(r) dr,
\end{equation}        
and $n_{\rm e}=(k_{\rm B}T/S)^{3/2}$.

Model V -- The same as Model IV except a varying metallicity 
$Z=0.3Z_{\odot}(t/t_0)$ is assumed 
instead of $Z=0.3Z_{\odot}$ for the rest four models, where $t_0$ denotes
the present cosmic epoch. The parameters of the five models are summarized
in Table 1.

 \begin{table*}
 \vskip 0.2truein
\caption{The parameters and legend for the models of gas distribution.}
 \vskip 0.2truein
 \begin{tabular}{cccccc}
 \tableline
 \tableline
Model   &  cooling & $n_{\rm e}$ & $T$  &    metallicity ($Z_{\odot}$)  & 
                  line-style\\
 \tableline
I       &    no    &  gas-traces-mass &  e.h.e.$^*$      &  $0.3$  & dot-dash\\
II      &    no    &  e.h.e.          &  isothermal      &  $0.3$  & dashed \\
III     &    yes   &  e.h.e.           &  isothermal      &  $0.3$  & dotted \\
IV      &    yes   &  gas-traces-mass &  e.h.e.          &  $0.3$  & solid \\
V       &    yes   &  gas-traces-mass &  e.h.e.          &  $0.3(t/t_0)$ &
                                                          dash dot dot dot\\
 \tableline
 \end{tabular}
 \parbox {8.5in}{$^{*}$Obtained by solving the equation of 
	            hydrostatic equilibrium.}
 \end{table*}

\section{Expectation for SZ cluster counts}

The total SZ flux observed at frequency $\nu$ 
by a cluster of mass $M$ at redshift $z$ is (e.g. Barbosa et al. 1996)
\begin{equation}
S_\nu (x,M,z) = \frac{g_{\nu}(x)}{D^2_a(z)}
	\left(\frac{\sigma _{\rm T}}{m_{\rm e}c^2}\right)
        \int k_{\rm B} T(r)n_{\rm e}(r) \;4\pi r^2dr, 
\label{snu0}
\end{equation}
where 
\begin{eqnarray}
g_{\nu}(x)&=&2\frac{(k_{\rm B}T_{\rm CMB})^3}{(h_{\rm p}c)^2}
\frac{x^4e^x}{(e^x-1)^2}
f_{\nu}(x);\\
f_{\nu}(x)&=&x\coth\frac{x}{2}-4,
\end{eqnarray}
$x=h_{\rm p}\nu/k_{\rm B}T_{\rm CMB}$, $T_{\rm CMB}=2.728$ K  
is the temperature of CMB (Fixsen et al. 1996),
 and $D_a$ is the angular diameter
distance to the cluster.  In the isothermal case, 
the integral in the right hand of equation (\ref{snu0}) 
can be replaced by the total mass of the cluster if the universal 
baryon fraction is introduced:
\begin{equation}
S_{\nu}=\frac{g_{\nu}(x)}{D_a^2(z)}
        \left(\frac{k_{\rm B}T}{m_{\rm e}c^2}\right)
	\left(\frac{f_b\sigma_{\rm T}}{\mu_{\rm e}m_{\rm p}}\right) \; M.
\label{snu1}
\end{equation}
This is the standard method adopted in the literature for the
theoretical prediction of SZ cluster counts (Model II).
While the assumption
of isothermality is more or less reasonable in terms of current
X-ray observations, the dependence of the gas fraction on temperature
claimed by many observations throws doubt upon the direct utilization
of equation (\ref{snu1}). Nevertheless, if the gas fraction $f_b$ is allowed to
vary according to mass or temperature, or the total gas mass is 
replaced by the hot gas component alone, $M_{\rm gas}(T)$,
we may modify the above equation to be 
\begin{equation}
S_{\nu}=\frac{g_{\nu}(x)}{D_a^2(z)}
        \left(\frac{k_{\rm B}T}{m_{\rm e}c^2}\right)
	\left(\frac{\sigma_{\rm T}}{\mu_{\rm e}m_{\rm p}}\right) 
	\; M_{\rm gas}(T).
\label{snu2}
\end{equation}
As shown in the above section, $M_{\rm gas}(T)$ can be analytically
determined for any cluster within the framework
of radiative cooling. This provides a simple approach to estimating the
effect of radiative cooling on SZ cluster counts without knowing
the density distribution of the hot, isothermal gas inside clusters, 
which will be applied to Model III. 
For rest three models, I, IV and V, we will adopt
the exact formula of $S_{\nu}$, equation (\ref{snu0}),
in the theoretical prediction of SZ cluster counts.

The expected number of SZ selected clusters with flux greater than
$S_{\nu}$ and in redshift interval ($z$, $z+dz$) is 
\begin{equation}
\frac{dN(> S_{\rm \nu})}{dzd \Omega} = \frac{dV}{dzd \Omega} \int _
{M_{\rm min}(z,S_{\rm \nu})} ^{\infty}  \frac{dn}{dM} dM,
\end{equation}
where the mass threshold $M_{\rm min}(z,S_{\rm \nu})$ is given by
the definition of equation (\ref{snu0}) or (\ref{snu1}) or 
(\ref{snu2}), depending on what 
approximation we would use. 
We adopt the PS  mass function to
describe the distribution of clusters
\begin{equation}
dn=-\sqrt{\frac{2}{\pi}} \frac{\bar{\rho}}{M} 
    \frac{\delta_{\rm c}(z)}{\sigma ^2} 
    \frac{d\sigma}{dM} 
    \exp{\left(-\frac{\delta_c^2(z)}{2\sigma^2} \right)} dM,
  \label{eq:ps}  
\end{equation} 
where $\bar{\rho}$ is the mean cosmic density, $\delta_c$ is the 
linear over-density of perturbations that collapsed and virialized at 
redshift $z$, and $\sigma$ is the linear theory variance of the mass density
fluctuation in sphere of mass $M$:
\begin{equation}
\sigma^2(M)=\frac{1}{2\pi^2} 
	\int \limits_0^{\infty} k^2 P(k) {\vert W(kR)\vert}^2 dk,
\end{equation}
and $W(kR)=3(\sin x-x\cos x)/x^3$  
is the Fourier representation of the window function.
The power spectrum, $P(k)\propto k^nT^2(k)$, is normalized by 
the rms fluctuation on an $8$ $h^{-1}$ Mpc scale, $\sigma_8$, and
we take the transfer function $T(k)$ from an adiabatic CDM model given
by Bardeen et al. (1986) for the Harrison-Zel'dovich case n=1.
Note that our results may be moderately changed if   
a modified mass function of dark halos is adopted
(e.g. Jenkins et al. 2001; Sheth \& Tormen 2001).

We work with a flat cosmological model ($\Lambda$CDM) of 
$\Omega_bh^2=0.019$, $\Omega_{\rm M}=0.3$ and $\Omega_{\Lambda}=0.7$.
We adopt $\sigma_8=0.90$ and a Hubble constant of $h=0.65$. 
We choose the SZ flux limit
to be $S_{\nu}=15$ mJy at frequency $\nu=30$ GHz. We perform numerical
calculations for the five models of gas distribution with and 
without cooling, and demonstrate the resulting differential SZ cluster counts 
in Figure 1.
While there are some differences between the redshift distribution of 
clusters predicted by the gas-traces-mass assumption and the one 
by the isothermal assumption, the effect of radiative cooling leads
to a moderate decrease in the expected number of SZ clusters 
although the peak locations with and without cooling remain roughly  
the same. Alternatively, our prediction alters
only slightly if the cosmic evolution of metallicity according to
$0.3Z_{\odot}(t/t_0)$ is included.

Recall that in the standard treatment, the expectation for SZ cluster 
counts is made by presumably taking the intracluster gas to be isothermal
and without cooling. 
If this prediction is directly applied to the determination
of cosmological parameters in future SZ surveys,
large uncertainty may be introduced because of the ignorance of
radiative cooling correction. Now we can evaluate the uncertainty in 
the determination of cosmological parameters by comparing 
the theoretical predictions with and without radiative cooling.
To do this, we still work with a flat cosmological model of
$\Omega_{\rm M}+\Omega_{\Lambda}=1$, but allow $\Omega_{\rm M}$ 
to vary according to $\sigma_8$ until the predicted redshift 
distribution of SZ cluster counts by the standard model (II) matches
those by the cooling models III--V.

Figure 2 shows the 
resulting $\sigma_8$ versus $\Omega_{\rm M}$, in which the 
contours indicate $68\%$ joint confidence intervals on 
the two parameters, corresponding to $\Delta\chi^2=2.30$.  
Uncertainties in $\Omega_{\rm M}$ and $\sigma_{\rm 8}$ 
due to radiative cooling correction
can be easily demonstrated by noticing that the true
input values in the cooling models are 
$\Omega_{\rm M}=0.30$ and $\sigma_{\rm 8}=0.90$, 
while our best-fit results are 
$(\Omega_{\rm M},\sigma_{\rm 8})=
 (0.24^{+0.13}_{-0.11},0.89^{+0.08}_{-0.10})$,
$(0.23^{+0.11}_{-0.11},0.88^{+0.08}_{-0.10})$ and 
$(0.23^{+0.10}_{-0.11},0.89^{+0.07}_{-0.10})$ for the 
three cooling models III, IV, and V, respectively.
Namely, without the correction of radiative cooling
we might underestimate both $\Omega_{\rm M}$ and $\sigma_8$ parameters
for $\Lambda$CDM cosmological model, although within the uncertainties
of our 'observations', the best-fit parameters are still consistent 
with the input values.

\section{SZ Power spectrum}

The angular power spectrum of temperature fluctuation on the CMB
sky due to the SZ effect of clusters  can be separated  
into the Poisson term $C_l^{(P)}$ and clustering term $C_l^{(C)}$
(Cole \& Kaiser 1988; Komatsu \& Kitayama 1999):
\begin{eqnarray}
  \label{eq:Cp}
  C_l^{(P)} &=& f_{\nu}^2(x)\int_0^{z_{\rm dec}} dz \frac{dV}{dz}\\
            & &   \int_{M_{\rm min}}^{\infty} dM
                \frac{dn(M,z)}{dM}|y_l(M,z)|^2 ,
\end{eqnarray}
and
\begin{eqnarray}
  \label{eq:Cc}
  C_l^{(C)} &=& f_{\nu}^2(x)\int_0^{z_{\rm dec}} dz \frac{dV}{dz} 
                P(k=l/D_0,z) \nonumber \\
	    & &	
                \left [\int_{M_{\rm min}}^{\infty} dM
                \frac{dn(M,z)}{dM} b(M,z) y_l(M,z) \right]^2 ,
\end{eqnarray}
where $z_{\rm dec}\approx1000$ is the CMB photon decoupling redshift,
$D_0$ is the comoving distance to cluster of mass $M$ at $z$,
$y_l(M,z)$ is the Fourier transform of the Compton $y$-parameter
in the thermal SZ effect: $\Delta T/T_{\rm CMB}=f_{\nu}(x)y(\theta)$,
$f_{\nu}(x)$ reflects the spectral dependence, and 
$b(M,z)$ is the so-called bias parameter, for which we use
the analytic approximation of Mo \& White (1996).
In our numerical computation, the minimum cluster mass
is taken to be $M_{\rm min}=1\times 10^{13}M_\odot$, 
and our final results are unaffected by this choice if we focus 
on large-scale ($>$arcminute) fluctuations on the CMB sky.

We illustrate in Figure \ref{clpg_5_cbibima} 
the SZ power spectra with and without cooling for
the five models in Table 1, together with the primary CMB signal produced
by CMBFAST.  It appears that the power spectra predicted by
the gas-traces-mass and isothermal assumptions become indistinguishable,
and the total SZ power spectrum is dominated by the Poisson distribution
of groups and clusters.  Inclusion of radiative cooling 
leads to a significant decrease of SZ power spectrum, especially 
at small angular scales or high-$\ell$.  This is simply because 
cooling removes more efficiently the hot gas at the central regions 
of clusters and in low-mass groups, and therefore 
suppresses the SZ signal at small scales. 
This result is in good agreement with the one found by hydrodynamical 
simulations (e.g. da Silva et al. 2001).

\section{Discussion and conclusions}

As a natural process, radiative cooling plays an important role in
the formation of galaxies. In the central regions of clusters and
groups, the typical cooling time scale of the hot gas
is usually less than the age of the universe. 
It follows that a considerable
amount of the intracluster/intragroup gas must  have cooled out of the 
hot phase since the formation of clusters and groups 
unless the cooling efficiency can be suppressed by 
other non-gravitational heating processes such as energy feedback
from supernovae and AGN activity. Since cooling removes the hot
gas in groups and clusters, it meanwhile reduces the SZ signal, 
leading to a decease of the expected SZ cluster counts and power spectrum.
This scenario has been confirmed by numerical simulations (e.g. da Silva et al.
2001) and our calculations based on a simple analytical model.
If the excess entropy detected in the central cores of groups and
clusters can be attributed to radiative cooling (Voit \& Bryan 2001),
some conclusions drawn from a phenomenological preheating model based 
on the observed entropy distribution can be
equally applied to the cooling scenario. Indeed, it has been shown that
the preheating model also results in a decrease of SZ
cluster counts and power spectrum (e.g. Komatsu \& Kitayama 1999;
Holder \& Carlstrom 2001; da Silva et al. 2001).
Recall that current X-ray
observations alone are still unable to distinguish between preheating and
cooling models (e.g. Voit et al. 2002; Borgani et al. 2002).

It is well known that radiative cooling is a runaway 
process. Consequently, the expected cooled materials within
the framework of radiative cooling from both  hydrodynamical 
simulations and analytical approaches exceed the observed stellar mass
fraction in the local universe (Balogh et al. 2001; 
Wu \& Xue 2002b and references therein). Inclusion of star formation
and energy feedback, which should in principle resolve the 
so-called cosmic cooling crisis, may significantly reduce the 
effect of radiative cooling on the X-ray properties and SZ signal  
of clusters. Indeed, recent numerical simulations by White et al. (2002) 
have shown that as a consequence of the combined effect of energy 
injection by star formation and radiative cooling, 
the SZ power spectrum remains roughly unchanged.

Our simple analytic approach to estimating the effect of radiative cooling
on the non-targeted SZ cluster survey and SZ power spectrum yields a
result consistent with those found by hydrodynamical simulations
(e.g. da Silva et al. 2001). This analytic method
allows an effective evaluation of the significance of 
various factors in the gas cooling process. 
For example, we have shown that within the framework of radiative 
cooling, both SZ cluster counts and power spectrum are insensitive
to temperature variation. In other words, the isothermality hypothesis 
is a good approximation in the theoretical study of cluster SZ effect.
Also, one may choose to use the gas-traces-mass assumption, which provides
a reasonable description of the gas distribution outside the X-ray core.
Recall that unlike the X-ray emission which depends on $n_{\rm e}^2$,
the SZ signal is proportional to $n_{\rm e}$. So, the gas outside the
X-ray core makes an important contribution to SZ signal too.
Finally, in order to eliminate the cooling crisis in the present model 
and include another natural process, i.e., heating by supernovae,
in the evolution of hot gas, 
it is worth exploring how energy injection into 
the intracluster/intrgroup gas from star formation
is incorporated into cooling scenario in a simple analytic way.

\acknowledgments

This work was supported by the National Science Foundation of China,
and the Ministry of Science and Technology of China, under Grant
No. NKBRSF G19990754.

\clearpage

\begin{figure}
\plotone{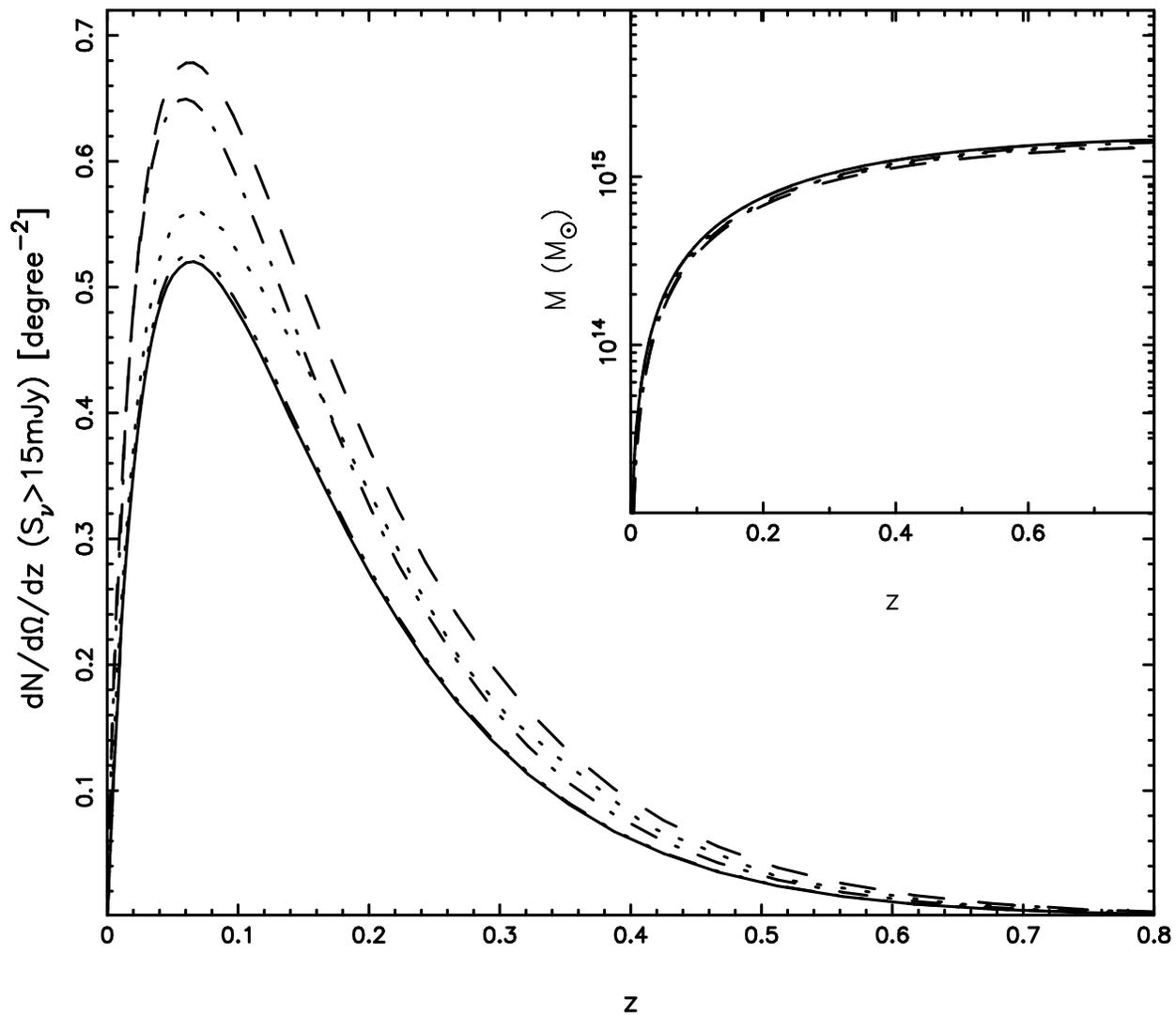}
\caption{Expected redshift distribution of SZ cluster counts with 
$S_{\rm \nu}=15 {\rm mJy}$ at frequency 30 GHz. The results for five models
with and without radiative cooling and under different assumptions
of the intracluster gas in Table 1 are shown. Also illustrated 
in the inset is the minimum mass threshold against cluster redshift. 
\label{szpg_5}}
\end{figure}

\clearpage 

\begin{figure}
\plotone{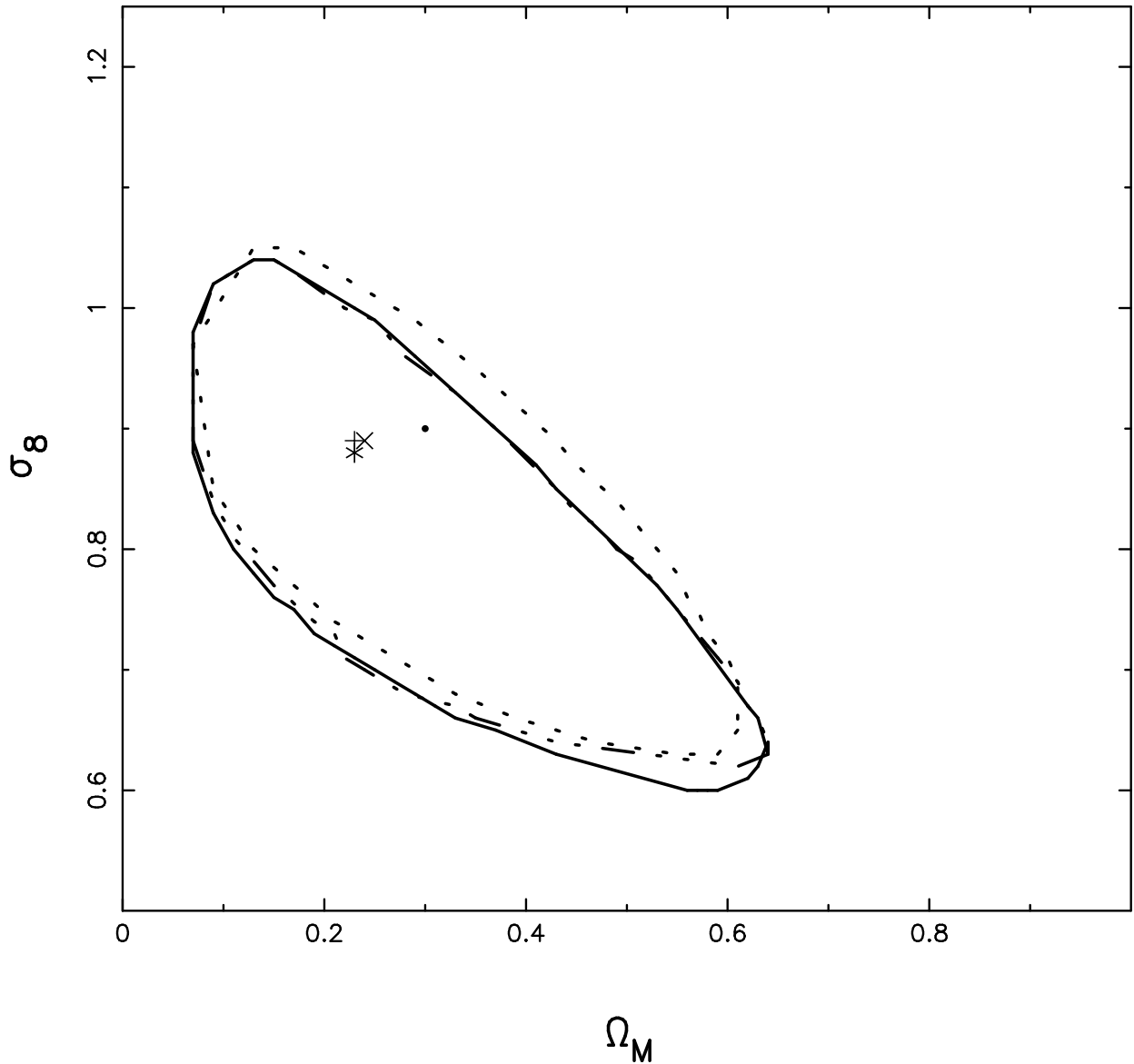}
\caption{$68\%$ joint confidence contours for $\Omega_{\rm M}$ and 
$\sigma_{\rm 8}$ obtained by minimizing the $\chi^2$ quantity:
The redshift distribution of SZ cluster counts predicted by standard 
model (II) is required to match those by cooling models (see Table 1 for 
legend). A flat cosmological model of 
$\Omega_{\rm M}+\Omega_{\Lambda}=1$ is assumed. For the 
cooling models we have adopted 
$(\Omega_{\rm M},\sigma_{\rm 8})=(0.30,0.90)$ 
(indicated by the filled circle).
The cross, asterisk  and plus symbols represent the best-fit results
for Model III, IV and V, respectively.
\label{68per_binpg}}
\end{figure}

\clearpage 

\begin{figure}
\plotone{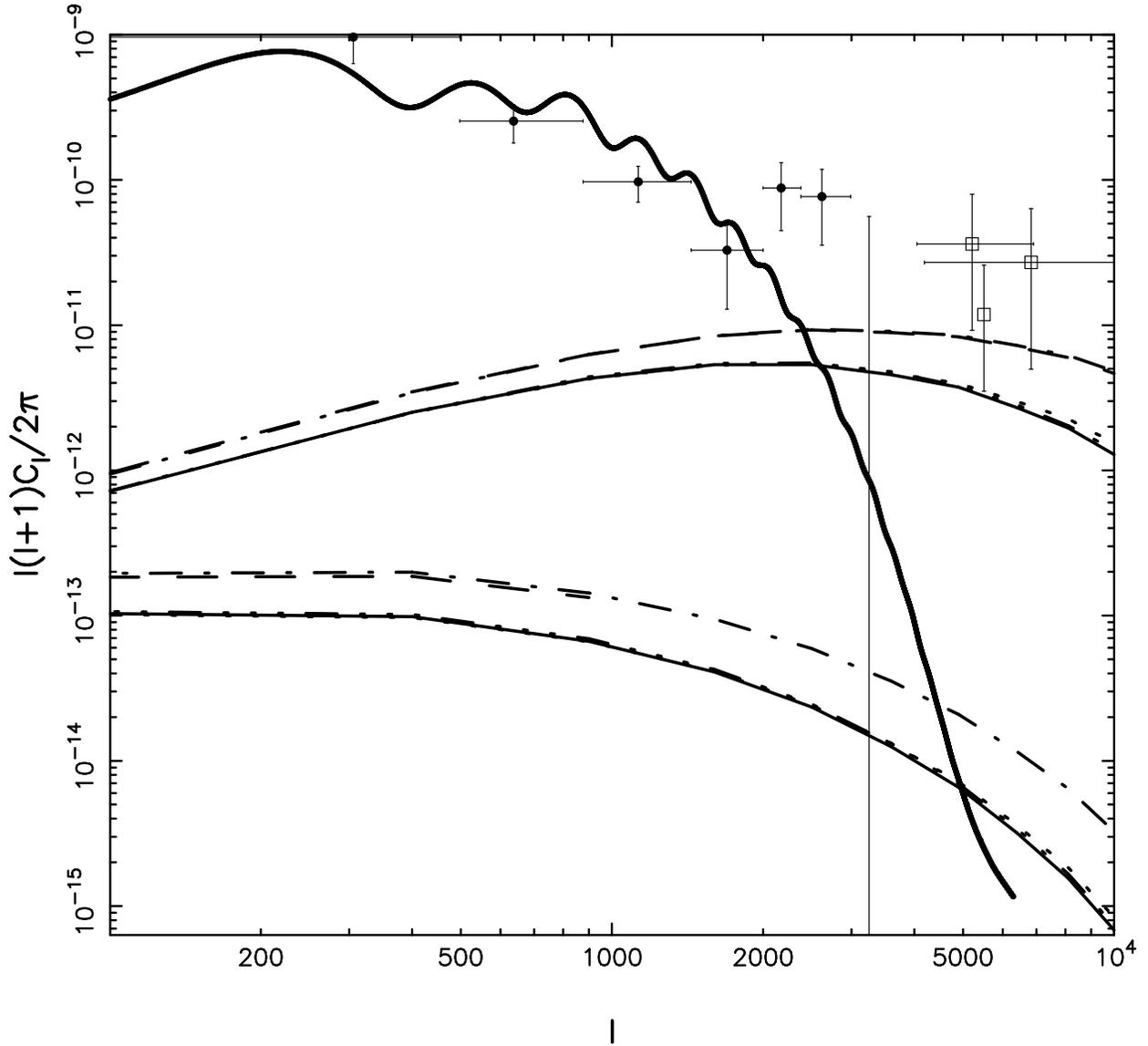}
\caption{SZ power spectra from intracluster gas  with and without
radiative cooling (see Table 1 for legend).
Top and bottom curves show the Poisson and clustering
contributions, respectively. The upper thick solid line is the primary
CMB signal predicted by CMBFAST. Recent observational results 
from BIMA (open squares; Dawson et al. 2001) and
CBI (filled circles; Mason et al. 2002) are also shown. 
\label{clpg_5_cbibima}}
\end{figure}

\end{document}